IDENTIFICATION AND STATISTICAL DECISION THEORY

Charles F. Manski
Department of Economics and Institute for Policy Research
Northwestern University

Abstract

Econometricians have usefully separated study of estimation into identification and statistical components. Identification analysis, which assumes knowledge of the probability distribution generating observable data, places an upper bound on what may be learned about population parameters of interest with finite sample data. Yet Wald's statistical decision theory studies decision making with sample data without reference to identification, indeed without reference to estimation. This paper asks if identification analysis is useful to statistical decision theory. The answer is positive, as it can yield an informative and tractable upper bound on the achievable finite-sample performance of decision criteria. The reasoning is simple when the decision-relevant parameter (true state of nature) is point identified. It is more delicate when the true state is partially identified and a decision must be made under ambiguity. Then the performance of some criteria, such as minimax regret, is enhanced by randomizing choice of an action. This may be accomplished by making choice a function of sample data. I find it useful to recast choice of a statistical decision function as selection of choice probabilities for the elements of the choice set. Using sample data to randomize choice conceptually differs from and is complementary to its traditional use to estimate population parameters.

I am grateful for the constructive comments of the editor and several reviewers.

1. Introduction

Econometricians have long found it instructive to separate the study of estimation into identification and statistical components. Koopmans (1949, p. 132) put it this way in the article that introduced the term *identification* into the literature: [1]

> "In our discussion we have used the phrase "a parameter that can be determined from a sufficient number of observations." We shall now define this concept more sharply, and give it the name *identifiability* of a parameter. Instead of reasoning, as before, from "a sufficiently large number of observations" we shall base our discussion on a hypothetical knowledge of the probability distribution of the observations, as defined more fully below. It is clear that exact knowledge of this probability distribution cannot be derived from any finite number of observations. Such knowledge is the limit approachable but not attainable by extended observation. By hypothesizing nevertheless the full availability of such knowledge, we obtain a clear separation between problems of statistical inference arising from the variability of finite samples, and problems of identification in which we explore the limits to which inference even from an infinite number of observations is suspect."

Koopmans recognized that statistical and identification problems limit in distinct ways the conclusions that may be drawn in empirical research. Statistical problems may be severe in small samples but diminish as the sampling process generates more observations. Identification problems cannot be solved by gathering more of the same kind of data. They can be alleviated only by invoking stronger assumptions or by initiating new sampling processes that yield different kinds of data.

---

[1] I adhere to Koopman's use of the term *identification* to consider "a hypothetical knowledge of the probability distribution of the observations," not the lesser knowledge yielded by a sample drawn from that distribution. Some recent econometric research studies inference with sample data using a concept that has been called 'weak identification,' but that actually concerns statistical inference. For example, Andrews and Mikusheva (2014) write (p. 195): "Weak identification commonly refers to the failure of classical asymptotics to provide a good approximation to the finite sample distribution of estimates and t-and Wald statistics in point-identified models where the data contains little information." Stock, Wright, and Yogo (2002) reviews this work.

Koopmans did not assert that separation of statistical inference and identification is mandatory. However, the research of a large community of econometricians has shown the separation to be highly useful in many contexts. A primary concern of early econometrics was estimation of simultaneous equations systems, which has been recast recently as estimation of treatment effects. Another primary concern has been estimation when data quality is imperfect due to missingness or mismeasurement. I find it difficult to imagine how our present understanding of these and other subjects would have developed without separation of statistical inference and identification.

Whereas study of identification has been central to econometrics, the concept does not appear in statistical decision theory as developed by Wald (1950). Indeed, estimation does not appear. Statistical decision theory shares with econometrics the concept of a true parameter value (called a state of nature) assumed to lie in a specified parameter space (a state space). As in econometrics, statistical decision theory supposes that one observes sample data drawn from a probability (sampling) distribution. However, statistical decision theory does not presume that a decision maker uses the sample data to estimate the true state of nature. Wald's core concept of a statistical decision function (SDF) embraces all mappings from sample data to an action, including ones that do not involve estimation of the true state of nature. Prominent decision criteria – maximin, minimax-regret, and maximization of subjective average welfare − do not perform estimation.

Wald formalized the performance of an SDF, which he termed *risk*, as an expected value across repeated samples. Risk is well-defined and potentially computable for any sample size and in any setting where the expectation exists. Wald did not study limit experiments of asymptotic statistical theory, where a finite sample size hypothetically increases toward infinity. A fortiori, he did not study the even more hypothetical setting of identification, where one is assumed to know the probability distribution generating observations.

Given that statistical decision functions need not perform estimation, it is natural to ask whether study of estimation is relevant to statistical decision theory. If so, one may further ask whether separation of estimation into statistical and identification components is useful when studying decision making with sample data. To paraphrase the final sentence of the passage from Koopmans (1949) quoted earlier, is it useful to separate problems of decision making arising from the variability of finite samples and problems of identification to which decision making even from an infinite number of observations is suspect?

The first question has a straightforward positive answer. Although SDFs need not perform estimation, they often do so. Applied economists commonly use SDFs having the form [data → estimation → action], first performing estimation and then using the estimate to choose an action. Koopmans and Hood (1953) posed this idea early on. Considering "The Purpose of Estimation," they focused on the use of estimates to make predictions and wrote that (p. 127): "estimates . . . can be regarded as raw materials, to be processed further into solutions of a wide variety of prediction problems."

Manski (2021) studied the prominent class of SDFs that perform *as-if optimization*, computing a point estimate of the true state of nature and then choosing an action that optimizes welfare as if the estimate is the true state. I also considered SDFs that use set-valued estimates, such as confidence sets or estimates of the identification regions of partially identified states of nature. As-if optimization is often computationally tractable, an important practical consideration. Theorems proving the consistency of point and set estimates suggest, albeit not prove, that as-if optimization performs well in decision making when sample size is sufficiently large.

This paper addresses the second question, which is somewhat subtle. The knowledge of the probability distribution of observations assumed in identification analysis is qualitatively more informative than any finite data sample. Knowledge of the probability distribution generating sample data enables one to shrink the state space, eliminating states of nature that are inconsistent with the known distribution. In econometric terminology, the knowledge presumed in identification analysis enables one to shrink the initial parameter space to an informative subset or a singleton, yielding partial or point identification respectively.

4The subtlety is that sample data ordinarily do not enable one to shrink the state space.[2] Wald's statistical decision theory shares this important attribute with standard estimation theory, in which one specifies a parameter space before data collection and does not alter it with observation of sample data. For example, theorems proving that estimators are consistent make assumptions that specify a fixed parameter space.[3]

I find that identification analysis can yield an informative upper bound on the achievable finite-sample performance of decision criteria. The reasoning is simple when the true state of nature is point identified. Knowledge of the true state of nature implies that the decision maker can choose an action that maximizes welfare. It is logically impossible to achieve more with sample data. The situation is more delicate if the true state is partially identified and the planner faces a problem of decision making under ambiguity; that is, no feasible action dominates all alternatives. Then a decision maker cannot maximize welfare. He can at most use some reasonable criterion to choose an action. The performance of some criteria for choice under ambiguity, such as minimax regret, is enhanced by permitting randomized choice of an action. The possibility of improvement in performance stems from the fact that randomization enables the planner to select any choice probabilities for actions, whereas non-randomized choice restricts the planner to select only among extreme values for choice probabilities.

---

[2] I write "ordinarily" because there are extraordinary exceptions. These occur if one observes sample data that cannot possibly occur in some state of nature. Then the data refute this state of nature.

One might be inclined to refute a state of nature if one observes sample data that are off the support of the probability distribution generating the data. This criterion for refutation seems reasonable intuitively. However, observation of data off the support is logically possible and hence does not definitively refute a state of nature.

Manski (2021) briefly discusses use of sample data to construct a confidence set for the unknown true state and performance of as-if optimization excluding states that do not lie within the confidence set. This yields SDFs that may be advantageous from the perspective of computational tractability, but they should be evaluated by their performance across the full state space. Shrinking the state space to contain only states within a confidence set is not justified in Wald's theory.

[3] Le Cam (1986) presented an asymptotic version of statistical decision theory in which some features of the feasible states of nature are presumed to shrink toward a particular state at a specified rate as sample size increases, the shrinkage process not depending on the particular data observed. Hirano and Porter (2020) review recent research in this tradition. Wald's theory does not contemplate shrinking any aspect of the state space with sample size.

Researchers often write abstractly that randomized choice may be accomplished by implementing a "randomization device," without specifying the device. In practice, randomized choice is generally performed by specifying a probability distribution on a sample space, drawing a realization from the distribution, and using the realization to choose an action. A familiar practice is to specify a uniform distribution on the unit interval, partition the interval into subsets associated with each action, draw a realization from the distribution, observe the subset within which the realization lies, and choose the action associated with that subset. Drawing from the distribution $U[0, 1]$ is not a necessity. Any distribution with sufficiently rich support will do.

Randomizing choice based on a realization from a uniform or other specified distribution formally employs sample data to make a decision. However, using sample data to randomize choice conceptually differs from and is complementary to the traditional use of sample data to estimate population parameters. When one uses a realization from a uniform or other specified distribution to randomize, the data are uninformative about the true state of nature and hence are useless for estimation. When one uses sample data to estimate the true state, the standard presumption is that the sampling distribution generating the data is not specified by the researcher but rather is a function of the true state, making the data informative about this state. See Section 2.2 for further discussion.

In principle, the Wald theory enables direct evaluation of the finite-sample performance of decision criteria. Hence one may ask why use of the theory may benefit from computing an informative upper bound on performance. A practical reason is that computation of the upper bound on performance delivered by identification analysis is often more tractable than computation of exact finite-sample performance, which is possible in principle but can be demanding in practice.

As prelude, Section 2 reviews basic statistical decision theory. Section 3 formalizes decision making with the knowledge of the probability distribution of observations assumed in identification analysis, but without accompanying sample data of any type, informative or uninformative. This simple extension of Section 2 yields the upper bound on finite-sample performance when the true state is point-identified.

Section 4 considers decision making when the true state of nature is partially identified, and one combines knowledge of the identification region with sample data. I find it useful to interpret choice of an SDF as selection of choice probabilities for the elements of the choice set. Section 5 presents easily implemented analytical findings for the important special case of choice between two actions. The concluding Section 6 calls for intensified study of *regret consistency* and other *criterion consistency* to further relate identification analysis to statistical decision theory.

## 2. Basic Statistical Decision Theory

This section reviews basic statistical decision theory, which makes no reference to identification. The presentation draws on Manski (2021).

### 2.1. Decisions without Sample Data

First consider decisions without sample data nor any other device to make randomized choice possible. A decision maker, called a planner for short, faces a choice set C and believes that the true state of nature lies in a specified state space S. An objective function $w(\cdot, \cdot): C \times S \rightarrow R^1$ maps actions and states into welfare. The planner wants to maximize true welfare, but he does not know the true state. Decision theorists have proposed various ways of using $w(\cdot, \cdot)$ to form functions of actions alone, which can be optimized. Throughout this paper I assume for simplicity that choice set C contains finitely many actions. I do not assume that S is finite, but I use max and min notation when taking extrema across S, without concern for subtleties that may make it necessary to use sup and inf operations.

One approach places a subjective probability distribution $\pi$ on S, computes average state-dependent welfare with respect to $\pi$, and maximizes subjective average welfare over C. The criterion solves

(1) $\quad \max_{c \in C} \int w(c, s) d\pi.$

Another approach seeks an action that, in some sense, works uniformly well over all of S. This yields the maximin and minimax-regret (MMR) criteria. These respectively solve the problems

(2) $\quad \max_{c \in C} \min_{s \in S} w(c, s).$

(3) $\quad \min_{c \in C} \max_{s \in S} [\max_{d \in C} w(d, s) - w(c, s)].$

In (3), $\max_{d \in C} w(d, s) - w(c, s)$ is the *regret* of action c in state s; that is, the degree of suboptimality.[4]

2.2. Statistical Decision Problems

Statistical decision problems suppose that the planner observes data generated by a sampling distribution which is a known one-to-one or many-to-one function of the state of nature. To express this, let the feasible sampling distributions be denoted $(Q_s, s \in S)$. Let $\Psi_s$ denote the sample space in state s; $\Psi_s$ is the set of samples that may be drawn under distribution $Q_s$. The literature typically assumes that the sample space does not vary with s and is known. I do likewise and denote the sample space as $\Psi$. A statistical decision function, $c(\cdot): \Psi \to C$, maps the sample data into a chosen action.

An SDF is a deterministic function after realization of the sample data, but it is a random function ex ante. Hence, an SDF generically makes a randomized choice of an action. The only exceptions are SDFs that make almost-surely data-invariant choices. An SDF $c(\cdot)$ is almost-surely data-invariant in state s if there exists a $d \in C$ such that $Q_s[c(\psi) = d] = 1$.

---

[4] I confine attention here to polar cases in which a planner asserts a complete subjective distribution on the state space, or none. A planner might assert a partial distribution, placing lower and upper probabilities on states as in Dempster (1968) or Walley (1991), and maximize minimum subjective average welfare or minimize maximum average regret. These criteria combine elements of averaging across states and concern with uniform performance across states. Some statistical decision theorists refer to them as Γ-maximin and Γ-minimax regret (Berger, 1985).

Given that SDFs are random functions, welfare using a specified SDF is a random variable ex ante. Wald's frequentist statistical decision theory evaluates the performance of SDF $c(\cdot)$ in state s by $Q_s\{w[c(\psi), s]\}$, the ex-ante distribution of welfare that it yields across realizations $\psi$ of the sampling process. In particular, Wald measured the performance of $c(\cdot)$ in state s by its expected welfare across samples; that is, $E_s\{w[c(\psi), s]\} \equiv \int w[c(\psi), s]dQ_s$.[5] Not knowing the true state, a planner evaluates $c(\cdot)$ by the state-dependent expected welfare vector $(E_s\{w[c(\psi), s]\}, s \in S)$, which is computable. This idea is applicable whenever the expected welfare vector exists and is finite.

Statistical decision theory has mainly studied the same decision criteria as has decision theory without sample data. Let $\Gamma$ denote the set of feasible SDFs, which map $\Psi \to C$. The statistical versions of criteria (1), (2), and (3) are

(4) $\quad \max_{c(\cdot) \in \Gamma} \int E_s\{w[c(\psi), s]\} \, d\pi,$

(5) $\quad \max_{c(\cdot) \in \Gamma} \min_{s \in S} E_s\{w[c(\psi), s]\},$

(6) $\quad \min_{c(\cdot) \in \Gamma} \max_{s \in S} \left( \max_{d \in C} w(d, s) - E_s\{w[c(\psi), s]\} \right).$

Subject to regularity conditions ensuring that the relevant expectations and extrema exist, problems (4) – (6) offer criteria for decision making with sample data that are broadly applicable in principle. The primary challenge is computational. Problems (4) − (6) have tractable analytical solutions only in certain cases. Computation commonly requires numerical methods to find approximate solutions.

---

[5] Wald formalized the objective of decision making as minimization of loss rather than maximization of welfare. He measured performance by *risk*, the negative of expected welfare.

As discussed earlier, observing sample data ordinarily do not enable one to shrink the state space. Nevertheless, sample data may enhance decision making through two mechanisms. First, SDFs randomize choice of an action, except when they are almost-surely data-invariant. Second, data are informative about the true state if the feasible sampling distributions ($Q_s$, s ∈ S) vary with s.

The scope for randomization is obvious when the sampling process includes an intentional randomizing device, such as generation of a realization $\psi_u$ of a U[0, 1] random variable. Then one may partition [0, 1] into specified sub-intervals $I_d$, d ∈ C of lengths $L_d$, d ∈ C and construct the SDF [$c(\psi_u) = d$ iff $\psi_u \in I_d$]. This SDF yields the choice probabilities $\{Q_s[c(\psi) = d] = L_d, d \in C\}$ for all s ∈ S.

This process for constructing SDFs gives full flexibility in randomizing choice, but it is useless from the perspective of estimation. The data realization $\psi_u$ is uninformative about the true state of nature, so the choice probabilities do not vary across states. The task of statistical decision theory is to develop SDFs that use informative data to enhance decision making.

2.2.1. Remarks on Representations of Randomized Choice

I observed above that an SDF is a random function ex ante and, hence, makes a randomized choice of an action. I have noted the important distinction between informative and uninformative sample data. I have not, however, formalized this distinction in my notation for sampling distributions and sample data, which I have denoted as $Q_s$ and $\psi$. To express the distinction, one may decompose $Q_s$ into a product form $Q_s = Q_{s,inf} \times Q_{rd}$ and, accordingly, decompose a sample data realization into $\psi = (\psi_{inf}, \psi_{rd})$. Here $Q_{s,inf}$ is the sampling distribution of informative (inf) data, which varies with the state of nature. $Q_{rd}$ is the sampling distribution used in an intentional randomizing device (rd), which is specified by the researcher and does not vary with the state of nature. Some readers may find this expanded notation helpful to keep it in mind, but it is unnecessary to the analysis of this paper. Hence, I use the compact ($Q_s$, $\psi$) notation throughout.

My formalization of randomized choice through the ($Q_s$, $\psi$) notation departs from the longstanding practice in statistical theory of representing intentional randomization as an auxiliary process layered on top of the randomization generated by collection of informative sample data. The practice has been to describe the decision maker as choosing a probability distribution over actions rather than a particular action. Selection of an action is said to be accomplished by using a randomizing device to draw an action from the chosen probability distribution. This description of intentionally randomized choice has been used in research on randomized tests and confidence intervals as well as in Wald's statistical decision theory.

In this paper I find that it simplifies notation and analysis to use the compact $(Q_s, \psi)$ notation to express the randomization produced by both informative sample data and the uninformative sample data generated by a randomizing device. Section 4.1 will show how this joint randomization process yields choice probabilities across actions. One might wonder whether there exist versions of intentional randomization that are not representable in the manner I do here, as generation of uninformative sample data. I am not aware of any.

3. Decisions with Knowledge of the Sampling Distribution

From here on, I relate identification analysis to statistical decision theory. This section considers decision making with the knowledge assumed in identification analysis, but without observation of sample data or any other means of randomizing choice of an action. Thus, supposing that some $s \in S$ is the true state of nature, I assume that the planner knows the probability distribution $Q_s$ that generates sample data but does not observe realizations drawn from $Q_s$.

Knowledge of $Q_s$ is useful to decision making to the extent that it shrinks the state space. Let S be the original state space of Section 2.1, without knowledge of $Q_s$. Let $S(Q_s) \subset S$ be the shrunken state space obtained with knowledge of $Q_s$. Thus, the planner knows that the true state s lies in $S(Q_s)$. The true state is point-identified if $S(Q_s) = s$. It is partially identified if $S(Q_s)$ is a proper non-singleton subset of S.

I consider decision making from two timing perspectives: ex post, after $Q_s$ is known, and ex ante, before it is known. Section 3.1 formalizes the ex-post problem, which is analogous to decision making

11without sample data. Section 2.1 described this problem with state space S. Now the planner chooses an action with knowledge of $S(Q_s)$.

Section 3.2 formalizes the ex-ante decision problem, which is analogous to Wald's study of decision making before sample data have been observed. Wald considered a planner who chooses an SDF $c(\cdot): \Psi \rightarrow C$, specifying the action that the planner would choose should any sample data $\psi$ be observed. Now the planner chooses a decision function $c(\cdot): (Q_s, s \in S) \rightarrow C$, specifying the action that the planner would choose should any sampling distribution $Q_s$ become known.

When performing ex ante analysis, it is useful to define *uniform* point identification. The true state is uniformly point-identified if the function mapping states into sampling distributions is one-to-one; then $S(Q_s) = s, \forall s \in S$. Manski (1988, Section 1.1) previously distinguished identification and uniform identification, with identification implicitly meaning point identification. See also the discussion in Molinari (2020, Section 3.1).

3.1. Ex Post Decisions

The new analogs of decision criteria (1) – (3) choose actions with knowledge of $S(Q_s)$ rather than S. The criteria are

(7) $\quad \max_{c \in C} \int w(c, s) d\pi[s|S(Q_s)]$ .

(8) $\quad \max_{c \in C} \min_{s \in S(Q_s)} w(c, s)$.

(9) $\quad \min_{c \in C} \max_{s \in S(Q_s)} [\max_{d \in C} w(d, s) - w(c, s)]$.

In (7), the subjective expectation is taken with respect to the posterior distribution $\pi[s|S(Q_s)]$, which truncates the prior distribution to the subset $S(Q_s)$.

If the true state is point-identified, criteria (7) and (8) both reduce to the deterministic optimization problem $\max_{c \in C} w(c, s)$. Criterion (9) reduces to $\min_{c \in C} [\max_{d \in C} w(d, s) - w(c, s)]$, which is equivalent to the problem $\max_{c \in C} w(c, s)$. Thus, with all three criteria, point identification enables the planner to maximize welfare.

The above shows that, when the true state is point-identified as s, the welfare performance of decisions made with sample data is bounded above by $\max_{c \in C} w(c, s)$. This maximization problem is commonly much simpler to solve than criteria (4) – (6). Thus, identification analysis supports statistical decision theory by yielding an informative and tractable upper bound on the performance of decision making with finite-sample data.

3.2. Ex-Ante Decisions

The new analogs of decision criteria (4) – (6) choose decision functions $c(\cdot): (Q_s, s \in S) \to C$ that select actions for all potential sampling distributions. The feasible decision functions, now labelled $\Gamma^*$, map $(Q_s, s \in S) \to C$. The criteria are

(10) $\quad \max_{c(\cdot) \in \Gamma^*} \int w[c(Q_s), s] d\pi,$

(11) $\quad \max_{c(\cdot) \in \Gamma^*} \min_{s \in S} w[c(Q_s), s],$

(12) $\quad \min_{c(\cdot) \in \Gamma^*} \max_{s \in S} \{ \max_{d \in C} w(d, s) - w[c(Q_s), s] \}.$

If the true state is uniformly point identified, knowledge of $Q_s$ implies knowledge of s. Letting $\Gamma^{**}$ be all decision functions $c(\cdot): S \to C$, the above criteria reduce to

$$(10') \qquad \max_{c(\cdot) \in \Gamma^{**}} \int w[c(s), s] \, d\pi,$$

$$(11') \qquad \max_{c(\cdot) \in \Gamma^{**}} \min_{s \in S} w[c(s), s],$$

$$(12') \qquad \min_{c(\cdot) \in \Gamma^{**}} \max_{s \in S} \left\{ \max_{d \in C} w(d, s) - w[c(s), s] \right\}.$$

Selecting the decision function to be $c^*(s) = \mathrm{argmax}_{d \in C}\, w(d, s),\ \forall\ s \in S$ solves each of problems (10') − (12'). Thus, with all three criteria, point identification enables the planner to maximize welfare.

### 4. Decisions with Knowledge of the Sampling Distribution and with Sample Data

Given knowledge of $Q_s$, with its implied knowledge of $S(Q_s)$, one might conjecture that there is no point in observing sample data drawn from $Q_s$. After all, with $S(Q_s)$ known, observation of sample data yields no further information about the true state. Section 3 showed that the conjecture is correct if the true state is point-identified.

The situation differs if the true state is partially identified and the planner faces a problem of decision making under ambiguity. Depending on the welfare function, the state space, and the decision criterion, randomized choice of an action may outperform any singleton choice. Sample data enable randomized choice. Thus, decision making combining knowledge of $Q_s$ with sample data may outperform decision making using knowledge of $Q_s$ alone.

Sections 4.1 and 4.2 develop the basic reasoning, which recasts choice of an SDF as selection of choice probabilities for the elements of C. Section 5 examines the important special case in which C contains two feasible actions.

### 4.1. Decision Making as Selection of Choice Probabilities

As described in Section 2.2, Wald considered a planner who chooses a statistical decision function before observing sample data. The planner evaluates the performance of SDF $c(\cdot): \Psi \to C$ in state s by expected welfare across samples; $E_s\{w[c(\psi), s]\} \equiv \int w[c(\psi), s] dQ_s$.

Consider a planner who chooses an SDF after learning the sampling distribution $Q_s$, but before observing sample data $\psi$ drawn from $Q_s$. Then the relevant versions of the subjective-expected-welfare, maximin, and MMR decision criteria are

$$(13) \quad \max_{c(\cdot) \in \Gamma} \int E_s\{w[c(\psi), s]\} d\pi[s|S(Q_s)],$$

$$(14) \quad \max_{c(\cdot) \in \Gamma} \min_{s \in S(Q_s)} E_s\{w[c(\psi), s]\},$$

$$(15) \quad \min_{c(\cdot) \in \Gamma} \max_{s \in S(Q_s)} [\max_{d \in C} w(d, s) - E_s\{w[c(\psi), s]\}],$$

where the feasible SDFs $\Gamma$ map $\Psi \to C$.

When $S(Q_s)$ is a proper subset of S, the maximin value in (14) is weakly larger than in (5) because expected welfare is minimized over $S(Q_s)$ rather than the larger set S. The MMR value in (15) is weakly smaller than in (6) because regret is maximized over $S(Q_s)$ rather than S. Hence, study of maximin and MMR decision making with the information assumed in identification analysis bounds the best performance of maximin and MMR decisions achievable with only sample data.

The situation differs with maximization of subjective expected welfare. The maximum in (13) may be larger or smaller than in (4), depending on the welfare function, the prior distribution, and the posterior distribution. This difference in findings may appear odd, but it has a simple explanation. Whereas the maximin and MMR criteria measure performance uniformly across states of nature, subjective expected welfare averages across states. Uniform performance must weakly improve when one shrinks the set over which uniformity is demanded. Average performance can become better or worse, depending on what states are eliminated by shrinking the state space.

Criteria (13) – (15) can be written in an equivalent form that eases their comparison with the criteria of Section 3. Knowledge of $Q_s$ implies that, for any SDF $c(\cdot)$ and state s, the planner can evaluate the choice probabilities $\{Q_s[c(\psi) = d], d \in C\}$ with which $c(\cdot)$ selects alternative actions. The expected welfare of $c(\cdot)$ in state s is

$$(16) \quad E_s\{w[c(\psi), s]\} = \sum_{d \in C} Q_s[c(\psi) = d] \cdot w(d, s).$$

Observe that expected welfare varies with $c(\cdot)$ only through the choice probabilities. Thus, given knowledge of $Q_s$, evaluation of the performance of $c(\cdot)$ is equivalent to evaluation of the choice probabilities that it yields.

Now let us use (16) to rewrite criteria (13) – (15). Given knowledge of $Q_s$, evaluation of the performance of all feasible SDFs is equivalent to evaluation of the set $\Delta_\Gamma(Q_s) \equiv \{Q_s[c(\psi) = d], d \in C; c(\cdot) \in \Gamma\}$ of all feasible vectors of choice probabilities. Let $\delta(Q_s, d), d \in C$ denote a feasible vector of choice probabilities. Then (13) – (15) are equivalent to the criteria

$$(13') \quad \max_{\delta(Q_s, \cdot) \in \Delta_\Gamma(Q_s)} \int \left[\sum_{d \in C} \delta(Q_s, d) \cdot w(d, s)\right] d\pi[s|S(Q_s)],$$

$$(14') \quad \max_{\delta(Q_s, \cdot) \in \Delta_\Gamma(Q_s)} \min_{s \in S(Q_s)} \sum_{d \in C} \delta(Q_s, d) \cdot w(d, s),$$

$$(15') \quad \min_{\Delta(Q_s, \cdot) \in \Delta_\Gamma(Q_s)} \max_{s \in S(Q_s)} \left[\max_{d \in C} w(d, s) - \sum_{d \in C} \delta(Q_s, d) \cdot w(d, s)\right].$$

Comparison of these decision criteria with (7)−(9) shows that combining knowledge of $Q_s$ with observation of sample data weakly improves the performance of decision making. In (7)−(9), the only feasible choice probabilities $\delta(Q_s, \cdot)$ were the vertices of $\Delta_\Gamma(Q_s)$, each placing probability one on a single action. Now interior choice probabilities are feasible as well.

Expansion of the feasible choice probabilities beyond the vertices is inconsequential when the criterion is to maximize subjective expected welfare. For each $d \in C$, the vector of polar choice probabilities $[\delta(Q_s, d) = 1; \delta(Q_s, d') = 0, d' \neq d]$ solves problem (13') if $\int w(d, s)d\pi[s|S(Q_s)] \geq \int w(d', s)d\pi[s|S(Q_s)]$, $d' \neq d$, and uniquely so if all inequalities are strict. Thus, randomized choice is never necessary and is generically sub-optimal.

The situation differs with the maximin and MMR criteria. In these cases, interior choice probabilities may yield better performance than any vertex of $\Delta_\Gamma(Q_s)$. Section 5 gives analytical findings when choice set C contains two actions.

If the sample space and sampling distribution are sufficiently rich, $\Delta_\Gamma(Q_s)$ is the entire |C|-dimensional unit simplex. The richness condition is easily satisfied. It suffices that $\Psi$ contain an interval on the real line and that $Q_s$ have positive density on this interval. Then the planner can select an SDF to yield any vector of choice probabilities. The sampling process yielding informative sample data may not per se satisfy these properties, perhaps because the sample space $\Psi$ is discrete. If so, the richness condition can be satisfied by augmenting the informative sampling process with an intentional randomizing device, such as one generating a realization from a U[0, 1] distribution.

4.2. Ex Ante Decisions

Section 4.1 considered a planner who makes a decision after learning $Q_s$, as in Section 3.1. A planner might behave in an ex ante manner, as in Section 3.2. He would then select vectors of choice probabilities for all sampling distributions ($Q_s$, $s \in S$) that may potentially be observed. The ex-ante versions of (13') – (15') are

(17) $$\max_{\delta(Q_s, \cdot) \in \Delta_\Gamma(Q_s), \, s \in S} \int \left[ \sum_{d \in C} \delta(Q_s, d) \cdot w(d, s) \right] d\pi(s),$$

(18) $$\max_{\delta(Q_s, \cdot) \in \Delta_\Gamma(Q_s), \, s \in S} \min_{s \in S} \sum_{d \in C} \delta(Q_s, d) \cdot w(d, s),$$

(19) $$\min_{\Delta(Q_s, \cdot) \in \Delta_\Gamma(Q_s), \, s \in S} \max_{s \in S} \left[ \max_{d \in C} w(d, s) - \sum_{d \in C} \delta(Q_s, d) \cdot w(d, s) \right].$$

With these ex ante criteria, the decision maker evaluates performance across the original state space S, cognizant that he will make a decision after learning the sampling distribution.

5. Decisions under Ambiguity with Binary Choice Sets

Section 4 showed that, with $Q_s$ known, randomizing choice may be beneficial if the true state is partially identified and the planner chooses under ambiguity. I consider here the important special case where choice set C contains two actions, say {a, b} and the welfare values {w(a, s), w(b, s), s ∈ S($Q_s$)} have bounded range. Let the lower and upper extreme values of the bounded welfares w(a, s) and w(b, s) across S($Q_s$) be denoted $\alpha_L(Q_s) \equiv \min_{s \in S(Q_s)} w(a, s)$, $\beta_L(Q_s) \equiv \min_{s \in S(Q_s)} w(b, s)$, $\alpha_U(Q_s) \equiv \max_{s \in S(Q_s)} w(a, s)$, and $\beta_U(Q_s) \equiv \max_{s \in S(Q_s)} w(b, s)$. The planner faces ambiguity if both actions are undominated; that is, if w(a, s) > w(b, s) for some values of s and w(a, s) < w(b, s) for other values.

I assume that all choice probabilities are feasible; thus, $\Delta_\Gamma(Q_s)$ is the entire unit simplex in $R^2$. Noting that $\delta(Q_s, a) + \delta(Q_s, b) = 1$, I define $\delta(Q_s) \equiv \delta(Q_s, b)$, implying that $\delta(Q_s, a) = 1 - \delta(Q_s)$. Then the planner's problem is to choose a value $\delta(Q_s) \in [0, 1]$. Criteria (13') – (15') reduce to

(20) $$\max_{\delta(Q_s) \in [0, 1]} [1 - \delta(Q_s)] \cdot \int w(a, s) \, d\pi[s|S(Q_s)] + \delta(Q_s) \cdot \int w(b, s) d\pi[s|S(Q_s)],$$

(21) $$\max_{\delta(Q_s) \in [0, 1]} \min_{s \in S(Q_s)} [1 - \delta(Q_s)] \cdot w(a, s) + \delta(Q_s) \cdot w(b, s),$$

(22)  $\min_{\delta(Q_s) \in [0, 1]} \max_{s \in S(Q_s)} \max \{w(a, s), w(b, s)\} - [1 - \delta(Q_s)] \cdot w(a, s) - \delta(Q_s) \cdot w(b, s).$

Manski (2007a, chapter 11; 2009) studied problems (20) – (22) in a different substantive context, where the planner assigns a treatment to each member of a large population of observationally identical persons and $\delta(Q_s)$ is the fraction of persons assigned to treatment b. In that context, choosing $0 < \delta(Q_s) < 1$ means treatment diversification rather than randomized choice of an action. The mathematical problem is the same for both interpretations of $\delta(Q_s)$.

The findings depend on the decision criterion used. It has already been shown that a planner who maximizes subjective expected welfare generically does not randomize choice. A planner who uses the maximin criterion randomizes when facing some state spaces but not others. One using the MMR criterion always randomizes. I summarize the maximin and minimax-regret analysis and findings below.

5.1. Maximin Decisions

To solve maximin problem (21) with $Q_s$ known, one first computes the minimum welfare attained by each value of $\delta(Q_s)$ across all feasible states $s \in S(Q_s)$. One then chooses $\delta(Q_s)$ to maximize this minimum welfare.

The maximin solution is simple if $[\alpha_L(Q_s), \beta_L(Q_s)]$ is a feasible value of $\{w(a, s), w(b, s)\}$. Then the unique maximin solution is $\delta(Q_s) = 0$ if $\alpha_L(Q_s) > \beta_L(Q_s)$ and $\delta(Q_s) = 1$ if $\alpha_L(Q_s) < \beta_L(Q_s)$. All $\delta(Q_s) \in [0, 1]$ are maximin solutions when $\alpha_L(Q_s) = \beta_L(Q_s)$. Thus, randomized choice is generically sub-optimal when $[\alpha_L(Q_s), \beta_L(Q_s)]$ is feasible. The maximin value of welfare with $Q_s$ known is max $[\alpha_L(Q_s), \beta_L(Q_s)]$.

Maximin welfare with $Q_s$ unknown is max $[\alpha_L(S), \beta_L(S)]$, where $\alpha_L(s) \equiv \min_{s \in S} w(a, s)$ and $\beta_L(Q_s) \equiv \min_{s \in S} w(b, s)$. Thus, the maximin value of learning that the sampling distribution is $Q_s$ is

(23) $$\max [\alpha_L(Q_s), \beta_L(Q_s)] - \max [\alpha_L(S), \beta_L(S)],$$

which is necessarily non-negative. Considered before data collection, when $Q_s$ is not yet known, the maximin value of learning the sampling distribution is at least the minimum of (23) over $s \in S$ and no greater than the maximum of (23) over $s \in S$.

Maximin decisions may randomize choice if $[\alpha_L(Q_s), \beta_L(Q_s)]$ is not feasible. Manski (2007, Chapter 11) gives a simple example. Let $S(Q_s) = \{s_0, s_1\}$ and let $\{w(a, s_0) = 1, w(b, s_0) = 0, w(a, s_1) = 0, w(b, s_1) = 1\}$. Then $[1 - \delta(Q_s)] \cdot w(a, s_0) + \delta(Q_s) \cdot w(b, s_0) = 1 - \delta(Q_s)$ and $[1 - \delta(Q_s)] \cdot w(a, s_1) + \delta(Q_s) \cdot w(b, s_1) = \delta(Q_s)$. Setting $\delta(Q_s) = \tfrac{1}{2}$ maximizes minimum expected welfare.

5.2. MMR Decisions

The MMR solution always randomizes with binary choice sets when the planner faces ambiguity. Let $S(Q_s)_a$ and $S(Q_s)_b$ be the subsets of $S(Q_s)$ on which actions a and b are superior. That is, $S(Q_s)_a \equiv \{s \in S(Q_s): w(a, s) > w(b, s)\}$ and $S(Q_s)_b \equiv \{s \in S(Q_s): w(b, s) > w(a, s)\}$. Both subsets are non-empty when the planner faces ambiguity. Let $M(Q_s)_a \equiv \max_{s \in S(Q_s)_a} [w(a, s) - w(b, s)]$ and $M(Q_s)_b \equiv \max_{s \in S(Q_s)_b} [w(b, s) - w(a, s)]$ be maximum regret on $S(Q_s)_a$ and $S(Q_s)_b$ respectively. Both $M(Q_s)_a$ and $M(Q_s)_b$ are positive under ambiguity.

Manski (2007a, Complement 11A) proves that the MMR value of $\delta(Q_s)$ is always interior to $[0, 1]$ in settings with ambiguity. The result is

(24) $$\delta(Q_s)_{MR} = \frac{M(Q_s)_b}{M(Q_s)_a + M(Q_s)_b}.$$

The minimax value of regret with $Q_s$ known is

$$(25) \quad M(Q_s)_a \cdot \delta(Q_s)_{MR} = M(Q_s)_a M(Q_s)_b / [M(Q_s)_a + M(Q_s)_b].$$

Expressions $M(Q_s)_a$ and $M(Q_s)_b$ simplify when $[\alpha_L(Q_s), \beta_U(Q_s)]$ and $[\alpha_U(Q_s), \beta_L(Q_s)]$ are feasible values of $[w(a, s), w(b, s)]$. Then $M(Q_s)_a = \alpha_U(Q_s) - \beta_L(Q_s)$ and $M(Q_s)_b = \beta_U(Q_s) - \alpha_L(Q_s)$. Hence, the MMR solution for choice under ambiguity is

$$(26) \quad \delta(Q_s)_{MR} = \frac{\beta_U(Q_s) - \alpha_L(Q_s)}{[\alpha_U(Q_s) - \beta_L(Q_s)] + [\beta_U(Q_s) - \alpha_L(Q_s)]}.$$

By analogous reasoning, minimax regret with $Q_s$ unknown is $M(S_a)M(S_b)/[M(S_a) + M(S_b)]$, where $S_a$ and $S_b$ are the subsets of $S$ on which actions a and b are superior; that is, $M(S_a) \equiv \max_{s \in S_a} [w(a, s) - w(b, s)]$ and $M(S_b) \equiv \max_{s \in S_b} [w(b, s) - w(a, s)]$. Thus, the minimax-regret value of learning that the sampling distribution is $Q_s$ is

$$(27) \quad M(Q_s)_a M(Q_s)_b / [M(Q_s)_a + M(Q_s)_b] - M(S_a)M(S_b)/[M(S_a) + M(S_b)],$$

which is necessarily non-positive. Considered before data collection, when $Q_s$ is not yet known, learning the sampling distribution reduces minimax regret by at least the maximum and no more than the minimum of (27) over $s \in S$.

It is interesting to compare the minimax value of regret (25) with the minimax regret that results if randomized choice is not possible, so the planner is only able to choose between the extreme choice probabilities 0 and 1. The solution then is $\delta(Q_s) = 0$ if $M(Q_s)_a \geq M(Q_s)_b$ and $\delta(Q_s) = 1$ if $M(Q_s)_a \leq M(Q_s)_b$. Maximum regret is $\min[M(Q_s)_a, M(Q_s)_b]$, which is larger than the minimax value of regret when randomized choice is permitted.

6. Towards Study of Criterion Consistency

Econometricians have usefully separated the study of estimation into identification and statistical components. Asymptotic theory has been used to connect identification to statistical inference. Consistency theorems show that, in the presence of regularity conditions, the hypothetical knowledge of the probability distribution of observations assumed in identification analysis is increasingly well-approximated as sample size increases.

This paper has shown that identification analysis can be useful to statistical decision theory, placing an informative and tractable upper bound on the performance of decision making with sample data. The argument is simple when the true state of nature is point identified, when it suffices to consider non-randomized choice among actions. When the true state is partially identified and a decision must be made under ambiguity, the performance of some criteria, such as minimax regret, is enhanced by permitting randomized choice of an action.

An important question not studied in this paper asks whether the upper bound on performance of finite-sample decision making delivered by identification analysis is attained asymptotically as sample size increases. The answer to this question may vary with the decision criterion as well as with the welfare function, state space, and sampling process. Hence, I will speak of *criterion consistency,* specifying the criterion under study.

Study of consistency of Bayes estimates in settings with point identification has a long history. Study of Bayesian estimation of partially identified parameters has begun to develop (e.g., Moon and Schorfheide, 2012; Kline and Tamer, 2016; Giacomini and Kitagawa, 2021). However, the present concern is the asymptotic performance of Bayesian decision making, not estimation. This question requires new analysis.

Considering regret, I posed the question of regret consistency in Manski (2004, footnotes 7 and 14), considering a commensurate sequence of sampling processes and SDFs indexed by sample size N. I defined this sequence to be *pointwise regret consistent* if regret converges to zero in all states of nature and uniformly regret consistent if maximum regret converges to zero. I showed that empirical success treatment rules are uniformly regret consistent when sample data on treatment response are generated by a classical randomized experiment and the covariates on which treatment choices are conditioned have finite support. On the other hand, Stoye (2009) showed that the minimax-regret treatment rule is data-invariant, hence not regret consistent, when covariates have continuous support and the state space does not restrict the variation of conditional mean treatment response across covariate values.

There is much scope to generalize research on regret consistency, considering other settings with point or partial identification. When the state of nature is point identified, the definitions of regret consistency given above should suffice. When it is partially identified and decisions must be made under ambiguity, maximum regret with the knowledge of the sampling distribution presumed in identification analysis is positive rather than zero. Hence, the definition of regret consistency must be generalized to call for convergence of finite-sample maximum regret to whatever maximum regret is achievable with knowledge of the sampling distribution.

Manski (2007b, 2021) studied a simple setting with severe partial identification, namely binary treatment choice using observational data with no assumptions on treatment selection. The analysis proved that, even with small positive sample size, analog estimation of the identification region and use of the estimate to randomize choice as in (26) yields finite-sample maximum regret equal to that achievable with knowledge of the sampling distribution, implying regret consistency. However, this is a very special case. Study of regret consistency more generally is complex in settings with partial identification. Stoye (2012) provides some analysis indicating the subtlety of the matter.

Whereas it appears reasonable to expect that Bayesian and regret consistency is achievable in broad settings of interest, the same cannot be said about maximin consistency. This was recognized verbally by Savage (1951), who stated that the maximin criterion is "ultrapessimistic" and wrote (p. 63): "it can lead to the absurd conclusion in some cases that no amount of relevant experimentation should deter the actor from behaving as though he were in complete ignorance." Formalizing this statement in a canonical case of randomized experimentation, Manski (2004) showed that the maximin criterion yields a data-invariant treatment rule in this context; hence, it is not maximin consistent.